\documentclass[runningheads]{llncs}
\usepackage{graphicx}
\usepackage[vlined,linesnumbered,ruled]{algorithm2e}
\usepackage{amsmath}
\usepackage{hyperref}
\hypersetup{
    colorlinks=true,
    linkcolor=blue,
    filecolor=magenta,      
    urlcolor=cyan,
}
\begin{document}
\title{CEDCES: A Cost Effective Deadline Constrained Evolutionary
Scheduler for Task Graphs in Multi-Cloud System}
\titlerunning{Cost Effective Deadline Constrained Scheduler}
%
\author{Atharva Tekawade  \and
 Suman Banerjee }
\authorrunning{Tekawade and Banerjee}
%
\institute{Department of Computer Science and Engineering, \\ Indian Institute of Technology Jammu, Jammu 181221, India.\\
\email{\{2018uee0137,suman.banerjee\}@iitjammu.ac.in}}
\maketitle              
\begin{abstract}
Many scientific workflows can be modeled as a Directed Acyclic Graph (henceforth mentioned as DAG) where the nodes represent individual tasks and the directed edges represent data and control flow dependency between two tasks. Due to large computational resource requirements, a single cloud cannot meet the requirements of the workflow. Hence, a multi-cloud system, where multiple cloud providers pool their resources together becomes a good solution. The major objectives considered while scheduling the tasks present in a task graph include execution cost and makespan. In this paper, we present \textbf{C}ost \textbf{E}ffective \textbf{D}eadline \textbf{C}onstrained \textbf{E}volutionary \textbf{S}cheduler (henceforth mentioned as CEDCES) which aims to minimize the execution cost under a given deadline constraint. CEDCES contains Particle Swarm Optimization-based (henceforth mentioned as PSO) method in its core, however includes novel initialization, crossover, and mutation schemes. Extensive simulation experiments on real-world workflows show that CEDCES outperforms the state-of-art algorithms, in particular, 60.41\% on average in terms of execution cost. In cases where the deadline is violated, CEDCES gives the least overshoot in execution time and outperforming the others by 10.96\% on average.
\end{abstract}
%
%
%
\section{Introduction}
Many real-world industrial, automotive, and avionic control systems can be represented as tasks and inter dependencies among them. They are often modelled as a directed acyclic graphs where each node represents a task and directed edge from one task to another task represents data and/or control flow relationship. Hence the second task can not be started unless the first one gets finished and its  output is available to the second one. These are called as task graphs or workflows \footnote{In this study both these terms have been used interchangeably.}. Due to the gigantic size of the workflows distributed computing platforms or multi cloud systems are used \cite{guo2018cost,roy2020contention}.

Scheduling the tasks of an workflow in a multi-cloud systems remains an active area of research \cite{tang2021reliability,guo2018cost}. This has been shown in the literature that this scheduling problem where the goal is to minimize the makespan is NP-hard \cite{ullman1975np}. Hence, several heuristic and meta heuristic solution approaches have been proposed \cite{pandey2010particle}. These studies are in diverse directions such as trust management in multi-cloud framework, in adaptive setting and many more. However, based on our literature review we conclude that existing studies on workflow scheduling mostly focuses on traditional distributed computing environments (such as grid), and only a few contributions are made in the cloud environment. Moreover, the existing works \cite{guo2018cost}, \cite{rodriguez2014deadline} do not consider the difference in pricing mechanisms across cloud providers. Hence there is enough scope to work in this area. In this paper, we study the problem of scheduling of tasks of an workflow in multi cloud system having different billing mechanism. In particular, we make the following contributions in this paper.
\begin{itemize}
   \item We model a fully interconnected multi-cloud system that models the different cloud providers and the resources they provide.
    \item We formulate and integrate the pricing mechanisms into our model including both computation and communication costs offered by the different cloud providers.
    \item We propose CEDCES, a PSO-based scheduling algorithm that minimizes execution cost under a given deadline constraint in a multi-cloud system. Further, we analyze the proposed methodology for its time requirement.
    \item We perform simulation experiments on real-world task graphs of different sizes and compare our method with the state-of-art approaches.
\end{itemize}

The rest of the paper is organized as follows. Section \ref{Sec:SM} describes the system's model and problem definition. The proposed solution has been described in Section \ref{Sec:PA}. Section \ref{Sec:Experiments} contains the experimental evaluation of the proposed solution methodology. Finally, Section \ref{Sec:CFA} concludes our study and gives future research directions.

\begin{algorithm}[H]
\caption{Maximum parallelizable tasks}
	\KwIn{Task graph}
    \KwOut{Set of tasks that can execute in parallel}
    $\mathcal{P} \longleftarrow \emptyset$\;
    Compute $top\_level(v)$ for each task using BFS\;
    Compute $max\_lvl$ as the topological level with most number of tasks\;
    Add all tasks having topological level as $max\_lvl$ in $\mathcal{P}$\;
    Order tasks based in increasing on decreasing order of $top\_level(v)$ in list $L$\;
    \For{$v_i \in L$} {
        \If{$top\_lvl(v_i) < max\_lvl$} {
            Add $v_i$ to $\mathcal{P}$ if it doesn't share dependency with any task already in $\mathcal{P}$\;
        }
    }
	\KwRet $\mathcal{P}$\;
	\label{Algo:1}
\end{algorithm}

\begin{algorithm}[h]
\caption{Particle to schedule mapping}
	\KwIn{Task graph, Multi-cloud system parameters, Particle ($X$)}
    \KwOut{Schedule}
    Compute resource pool $\mathcal{R}$ using Algorithm 1\;
    $\mathcal{M}, \mathcal{R}_{curr} \longleftarrow \emptyset$\;
    $TEC, TET \longleftarrow 0$\;
    \For{$v_i \in V$}{
        $vm_{u_i} \longleftarrow \mathcal{R}[X[v_i]]$\;
        $VM(k, p) \longleftarrow type(vm_{u_i})$\;
        $exec \longleftarrow \frac{w(v_i)}{w(VM(k, p))}$\;
        $ST_{v_i} \longleftarrow 0$\;
        \For{$v_j \in pred(v_i)$}{
            $ST_{v_i} \longleftarrow \max(ST_{v_i}, FT_{v_j})$\;
        }
        $transfer \longleftarrow 0$\;
        \For{$v_j \in succ(v_i)$}{
            $vm_{u_j} \longleftarrow \mathcal{R}[X[v_j]]$\;
            $VM(k', p') \longleftarrow type(vm_{u_j})$\;
            \If{$u_i \neq u_j$}{
                \If{$k \neq k'$}{
                    $transfer \longleftarrow transfer + \frac{w(v_i, v_j)}{B_{k, k'}}$\;
                     Find pricing $c_{k, k'}$ according to Table \ref{tab2}\;
                     $TEC \longleftarrow TEC + c_{k, k'} \cdot w(v_i, v_j)$\;
                }
                \Else{
                    $transfer \longleftarrow transfer + \frac{w(v_i, v_j)}{B_{k}}$\;
                }
            }
        }
        $PT_{v_i} \longleftarrow exec + transfer$\;
        \If{$vm_{u_i} \in \mathcal{R}_{curr}$}{
            $ST_{v_i} \longleftarrow \max(ST_{v_i}, LFT_{vm_{u_i}})$\;
        }
        \Else{
            $ST_{v_i} \longleftarrow \max(ST_{v_i}, T_{boot}[VM(k,p)])$\;
            $LST_{vm_{u_i}} \longleftarrow ST_{v_i} - T_{boot}[VM(k,p)]$\;
            $\mathcal{R}_{curr} \longleftarrow \mathcal{R}_{curr} \cup \{vm_{u_i}\}$\;
        }
        $FT_{v_i} \longleftarrow ST_{v_i} + PT_{v_i}$\;
        $LFT_{vm_{u_i}} \longleftarrow FT_{v_i}$\;
        $\mathcal{M} \longleftarrow \mathcal{M} \cup \{(v_i, vm_{u_i}, ST_{v_i}, FT_{v_i})\}$\;
    }
    $TET \longleftarrow FT_{v_n}$\;
    \For{$vm_r \in \mathcal{R}_{curr}$}{
        Compute $cost$ corresponding to lease period $LFT_{vm_r}-LST_{vm_r}$ using Equation No. \ref{eqn5}, \ref{eqn6}\;
        $TEC \longleftarrow TEC + cost$\;
    }
    $\mathcal{S} \longleftarrow (\mathcal{R}, \mathcal{M}, TEC, TET)$\;
	\KwRet $\mathcal{S}$\;
	
	\label{Algo:2}
\end{algorithm}

\section{System Model and Problem Formulation} \label{Sec:SM}
In this section, we describe the system's model and describe our problem formally. For any positive integer $n$, $[n]$ denotes the set $\{1, 2, \ldots, n\}$. Initially, we start by describing the task graph.

\subsection{Workflow}
A scientific workflow can be modeled as a Directed Acyclic Graph denoted by $G(V, E)$, where $V = \{v_i : i \in [n]\}$ is the set of vertices denoting the individual tasks, and $E(G) = \{(v_i, v_j) : v_i, v_j \in V(G)\}$ is the set of edges. An edge $(v_i, v_j)$ denotes a dependency between tasks $v_i$ and $v_j$ \emph{i.e.} $v_j$ cannot start unless $v_i$ finishes it's execution and transfers the required output completely to $v_j$. Each task node $v_i$ has a computational resource requirement denoted by $w(v_i)$. The weight of the edge $(v_iv_j)$ is denoted by $w(v_i, v_j)$ and it signifies the amount of output that needs to be transferred from $v_i$ to $v_j$. Below, we list some definitions that will be used subsequently.

\begin{definition} \label{def1}
$pred(v_i)$ denotes the set of immediate predecessor tasks of the task $v_i$. Mathematically, $pred(v_i) = \{v_j : (v_j, v_i) \in E \} $.
\end{definition}  
\begin{definition} \label{def2}
$succ(v_i)$ denotes the set of immediate successor tasks of the task node $v_i$. Mathematically, $succ(v_i) = \{v_j : (v_i, v_j) \in E \} $.
\end{definition}
\begin{definition} \label{def3}
$v_{entry}$ denotes the entry task. It is a redundant node having an outgoing edge with zero weight to every $v$ such that $pred(v) = \emptyset$. For simplicity we assume that $v_{entry} = v_1$.
\end{definition}
\begin{definition} \label{def4}
$v_{exit}$ denotes the exit task. It is a redundant node having an incoming edge with zero weight from every $v$ such that $succ(v) = \emptyset$. For simplicity we assume that $v_{exit} = v_n$.
\end{definition}
\begin{definition} \label{def5}
$top\_level(v)$ denotes the topological level of task node $v$, given by the below equation:

\begin{equation} \label{eqn1}
\scriptsize
top\_level(v) = \left\{ 
  \begin{array}{ c l }
   0 & \quad \textrm{if } v = v_1 \\
   \max_{u \in pred(v)} \{top\_level(u) + 1\} & \quad \textrm{otherwise}
  \end{array}
\right.
\end{equation}
The topological level can be computed using the Breadth-first search algorithm (BFS) \cite{bundy1984breadth}.
\end{definition}

\subsection{Multi-Cloud System}
Our model consists of m different cloud providers, each providing their own set of resources. These resources are offered in the form of Virtual Machines (henceforth mentioned as VMs). Assume that the $k^{th}$ cloud provider offers a total of $m_k$ different VM types. Let $VM(k, p)$ denote the $p^{th}$ type VM offered by the $k^{th}$ cloud provider. A VM is characterized by their CPU, Disk and Memory. We assume that VMs have sufficient memory to execute the
workflow tasks \cite{rodriguez2014deadline}. Let $w(VM(k, p))$ denote the processing capacity of $VM(k, p)$ (amount of computation performed in a second). The higher the processing capacity of a VM, the faster it executes a task. The amount of time it takes to execute $v_i$ on $VM(k, p)$ is denoted by $T_{exec}[v_i, VM(k, p)]$ and is given by Equation No. \ref{eqn2}. Additionally, we assume that a newly launched VM needs a specific initial boot time denoted by $T_{boot}[VM(k, p)]$.

\begin{equation} \label{eqn2}
T_{exec}[v_i, VM(k,p)] = \frac{w(v_i)}{w(VM(k, p))}
\end{equation}

\subsection{Network}
VMs belonging to the same cloud are connected by high speed internal network, whereas those belonging to different clouds are connected by slower external network \cite{tang2021reliability}. Let $B_k$ denotes the bandwidth of the communication links of the VMs within the $k$-th cloud. Similarly, let $B_{k, k'}$ denotes the bandwidth of the communication links connecting the $k$-th and $k^{'}$-th clouds. Communication time between two tasks $v_i$ and $v_j$ is denoted by $T_{comm}[v_i, v_j]$ and this depends on the amount of data to be transferred and the clouds where the tasks are hosted. This can be computed using Equation No. \ref{eqn3}.

\begin{equation} \label{eqn3}
\scriptsize
T_{comm}[v_i, v_j] = \left\{ 
  \begin{array}{ c l }
  0 & \quad \textrm{if } \text{$v_i, v_j$ are scheduled on the same VM instance} \\
  \frac{w(v_i, v_j)}{B_{k, k'}} & \quad \textrm{else if } k' \neq k \\
    \frac{w(v_i, v_j)}{B_k} & \quad \textrm{otherwise}
  \end{array}
\right.
\end{equation}
where, $v_i, v_j$ are assumed to be scheduled on the $k$-th and $k^{'}$-th clouds, respectively. \\

\subsection{Resource Model}
Due to the flexibility of resource acquisition provided by cloud providers, a client can run any number of instances of any type of VM. Let $\mathcal{R} = \{vm_r : r \in [\infty]\}$ denote the set of VM instances to be leased (pool of resources). Let $LST_{vm_r}$ and $LFT_{vm_r}$ denote the start and end times respectively for which $vm_r$ is leased. A VM needs to be kept on till a task has transferred data to all it's successor nodes \cite{rodriguez2014deadline}. Hence, the total processing time of a task $v_i$ denoted by $PT_{v_i}$ includes both execution and data-transfer as given by Equation No. \ref{eqn4}.

\begin{equation} \label{eqn4}
    \scriptsize
    PT_{v_i} = \left\{ 
      \begin{array}{ c l }
       T_{exec}[v_i, VM(k,p)] & \quad \textrm{if } v = v_n \\
       T_{exec}[v_i, VM(k,p)] + \underset{v_j \in succ(v_i)}{\sum} T_{comm}[v_i, v_j] & \quad \textrm{otherwise}
      \end{array}
    \right.
\end{equation}

where $v_i$ is assumed to be executed on VM of type $VM(k, p)$.

\begin{algorithm}[H]
\caption{VM initialization}
	\KwIn{Task graph, Multi-cloud system parameters, Deadline}
    \KwOut{Particle ($X$)}
    Compute random task-order $ord$ using topological sort\;
    Compute resource pool $\mathcal{R}$ using Algorithm 1\;
    Compute $MET$ values according to Equation No. \ref{eqn10}\;
    $\mathcal{R}_{curr} \longleftarrow \emptyset$\;
    \For{$v_i \in ord$}{
        \For{$vm_r \in \mathcal{R}$}{
            $VM(k, p) \longleftarrow type(vm_r)$\;
            $ST_{v_i} \longleftarrow 0$\;
            \For{$v_j \in pred(v_i)$} {
                Calculate $T_{comm}[v_j, v_i]$ using Equation No. \ref{eqn3}\;
                $ST_{v_i} \longleftarrow \max(ST_{v_i}, FT_{v_j} + T_{comm}[v_j, v_i])$\;
            }
            \If{$vm_r \in \mathcal{R}_{curr}$}{
                $ST_{v_i} \longleftarrow \max (ST_{v_i}, LFT_{vm_r})$\;
                \If{$ST_{v_i} + MET[v_i, VM(k, p)] \leq D$}{
                    $lease\_period \longleftarrow MET[v_i, VM(k, p)]$ \; 
                }
            }
            \Else{
                $ST_{v_i} \longleftarrow \max (ST_{v_i}, T_{boot}[VM(k, p)])$\;
                \If{$ST_{v_i} + MET[v_i, VM(k, p)] \leq D$}{
                    $lease\_period \longleftarrow T_{boot}[VM(k, p)] + MET[v_i, VM(k, p)]$\; 
                }
            }
            Compute $cost$ corresponding to $lease\_period$ using Equation No. \ref{eqn5}, \ref{eqn6}.\; 
        }
        Assign VM instance $vm_{u_i}$ satisfying deadline with minimum $cost$ to $v_i$\;
        If no VM instance satisfies the deadline, assign $vm_{u_i}$ with least deadline violation to $v_i$\;
        $VM(k, p) \longleftarrow type(vm_{u_i})$\;
        \For{$v_j \in pred(v_i)$}{
            $FT_{v_j} \longleftarrow FT_{v_j} + T_{comm}[v_j, v_i]$\;
            $u_j \longleftarrow X[v_j]$\;
            $LFT_{vm_{u_j}} \longleftarrow \max(LFT_{vm_{u_j}}, FT_{v_j})$\;
        }
        $ST_{v_i} \longleftarrow 0$\;
        \For{$v_j \in pred(v_i)$}{
            $ST_{v_i} \longleftarrow \max(ST_{v_i}, FT_{v_j})$\;
        }
        \If{$vm_{u_i} \notin \mathcal{R}_{curr}$}{
            $\mathcal{R}_{curr} \longleftarrow \mathcal{R}_{curr} \cup \{vm_{u_i}\}$\;
            $ST_{v_i} \longleftarrow \max (ST_{v_i}, LFT_{vm_{u_i}})$\;
        }
        $PT_{v_i} \longleftarrow T_{exec}[v_i, VM(k,p)]$\;
        $FT_{v_i} \longleftarrow ST_{v_i} + PT_{v_i}$\;
        $LFT_{vm_{u_i}} \longleftarrow FT_{v_i}$\;
        $X[v_i] \longleftarrow u_i$\;
    }
	\KwRet $X$\;
	\label{Algo:3}
\end{algorithm}

\begin{algorithm}[h]
\caption{CEDCES algorithm}
	\KwIn{Task graph, Multi-cloud system parameters, Deadline}
    \KwOut{Schedule}
    $ITR \longleftarrow$ Number of iterations\;
    $NUM \longleftarrow$ Number of particles\;
    \For{$i =$ 1 to $NUM$}{
        Initialize particle position $X_i^0$ according to Algorithm 2\;
        Initialize particle velocity $V_i^0$ randomly\;
        Set the personal best $pbest_i^0$ to $X_i^0$\;
        Update $gbest^0$ if $X_i^0$ is fitter\;
    }
    \For{$t =$ 1 to $ITR$}{
        \For{$i =$ 1 to $NUM$}{
            Compute $X_i^t, V_i^t$ according to Equation No. \ref{eqn7}, \ref{eqn8}\;
            Repair $X_i^t$ using SHR method if it is out of bounds\;
            Compute the schedule $\mathcal{S}$ of $X_i^t$ using Algorithm 3\;
            Update $pbest_i^t$ if $X_i^t$ is fitter\;
            Update $gbest^t$ if $X_i^t$ is fitter\;
        }
        Select $X_i^t, X_j^t$ using a binary selection tournament\;
        Apply single point crossover at random index $k$ on $X_i^t, X_j^t$ to get a new particle $\overline{X}^t$\;
        Replace the particle having the least fittest personal best with $\overline{X}^t$\;
        Apply mutation to a randomly chosen particle $X_i^t$ at random index $k$\;
    }
    Compute $\mathcal{S}$ for particle $gbest^{ITR}$ using Algorithm 3\;
	\KwRet $\mathcal{S}$\;
	\label{Algo:4}
\end{algorithm}

\subsection{Pricing Mechanisms}
In this study, we consider three popular cloud-providing services: Microsoft Azure (\emph{MA}), Amazon Web Services (\emph{AWS}), and Google Cloud Platform (\emph{GCP}). Each cloud provider charges the customer after a specified billing period $\tau$. Let $c_{k, p}$ denotes the price for renting $VM(k,p)$ for a single billing period. Below we discuss the pricing mechanisms used by the above-mentioned cloud providers from VM instance $vm_r$ \cite{tang2021reliability}.

\begin{itemize}
    \item \emph{MA} follows a fine-grained scheme where the customer is charged per minute of usage \emph{i.e.} $\tau = $ 1 min. The cost is given in Equation No. \ref{eqn5}.
    
    \begin{equation} \label{eqn5}
        cost = \lceil \frac{LFT_{vm_r} - LST_{vm_r}}{\tau} \rceil \cdot c_{k, p}
    \end{equation}
    
    \item \emph{AWS} follows a coarse-grained pricing mechanism where the customer per hour of usage. \emph{i.e.} $\tau = $ 1 hr. The cost can be obtained by Equation No. \ref{eqn5}.
    
    \item \emph{GCP} follows a hybrid pricing mechanism where the customer is charged for a minimum of ten minutes, after which per minute billing is followed. In this case, let the price for the first ten minutes be $C_{k,p}$ and $c_{k, p}$ as usual denotes he price per minute henceforth. The cost is formulated in Equation No. \ref{eqn6} where $\tau = $ 1 min.
    
    \begin{equation} \label{eqn6}
        cost = C_{k, p} + \max(0, \lceil \frac{LFT_{vm_r} - LST_{vm_r} - 10 \cdot \tau}{\tau} \rceil) \cdot c_{k, p}
    \end{equation}

\end{itemize}
Table \ref{tab1} illustrates the prices of various VMs \cite{tang2021reliability}. Apart from this, each cloud provider has a specific pricing scheme associated with sending data out. Let $c_{k, k'}$ denote the price per unit data for sending data between the $k^{th}$ and $k'^{th}$ providers. This price depends on many factors: Location where the VMs are hosted, size of data, etc. Table \ref{tab2} shows the rates for transferring data as taken from the official websites \cite{ma}, \cite{aws}, \cite{gcp}. In the table, across centers refers to locations managed by the same cloud provider but located in different places. Depending on the location (US, Europe, Asia, etc.) the prices vary. For simplicity, we consider the median value across all locations. Across clouds refers to the transfers across different cloud providers over external internet. Again this price depends on the location, so we consider the median value.

\begin{table*}
		\centering
		\begin{tabular}{|c|c|c|c|c|c|c|} 
			\hline
			\multicolumn{2}{|c|}{\emph{MA}} & \multicolumn{2}{c|}{\emph{AWS}} & \multicolumn{3}{c|}{\emph{GCP}} \\
			\hline
			\textbf{\emph{VM}} & \textbf{\emph{Per Minute(\$)}} & \textbf{\emph{VM}} & \textbf{\emph{Per hour(\$)}} & \textbf{\emph{VM}} & \textbf{\emph{Ten Minutes(\$)}} & \textbf{\emph{Per Minute(\$)}}\\
			\hline
			\textbf{B2MS} & 0.0015 & \textbf{m1.small} & 0.06 & \textbf{n1-highcpu-2} & 0.014 & 0.0012 \\
			\hline
			\textbf{B4MS} & 0.003 & \textbf{m1.medium} & 0.12 & \textbf{n1-highcpu-4} & 0.025 & 0.0023 \\
			\hline
			\textbf{B8MS} & 0.006 & \textbf{m1.large} & 0.24 & \textbf{n1-highcpu-8} & 0.05 & 0.0047 \\
			\hline
			\textbf{B16MS} & 0.012 & \textbf{m1.xlarge} & 0.45 & \textbf{n1-highcpu-16} & 0.1 & 0.0093 \\
			\hline
		\end{tabular}
		\vspace{0.1 cm}
		\caption{VM costs for different cloud providers}
	\label{tab1}
\end{table*}

\begin{table*}
		\centering
		\begin{tabular}{|c|c|c|c|c|c|c|} 
			\hline
			\multicolumn{2}{|c|}{\emph{MA}} & \multicolumn{2}{c|}{\emph{AWS}} & \multicolumn{2}{c|}{\emph{GCP}} \\
			\hline
			\multicolumn{2}{|c|}{\textbf{Same center - Free}} & \multicolumn{2}{c|}{\textbf{Same center- Free}} & \multicolumn{2}{c|}{\textbf{Same center - Free}} \\
			\hline
			\multicolumn{2}{|c|}{\textbf{Across centers - \$0.08/GB}} & \multicolumn{2}{c|}{\textbf{Across centers - \$0.02/GB}} & \multicolumn{2}{c|}{\textbf{Across centers - \$0.05/GB}} \\
			\hline
			\multicolumn{6}{|c|}{\textbf{Across clouds}} \\
			\hline
			\textbf{\emph{Data size}} & \textbf{\emph{Per GB(\$)}} & \textbf{\emph{Data size}} & \textbf{\emph{Per GB(\$)}} & \textbf{\emph{Data size}} & \textbf{\emph{Per GB(\$)}}\\
			\hline
			Upto 100GB & Free & Upto 100GB & Free & 0-1TB & 0.19 \\
			\hline
			First 10 TB & 0.11 & First 10 TB & 0.09 & 1-10TB & 0.18 \\
			\hline
			Next 40TB & 0.075 & Next 40TB & 0.085 & 10TB+ & 0.15 \\
			\hline
			Next 100TB & 0.07 & Next 100TB & 0.07 & \multicolumn{2}{|c|}{-} \\ 
			\hline
			Next 350TB & 0.06 & Greater than 150 TB & 0.05 & \multicolumn{2}{|c|}{-} \\ 
			\hline
		\end{tabular}
		\vspace{0.1 cm}
		\caption{Data-Transfer costs for different cloud providers}
	\label{tab2}
\end{table*}

\subsection{Problem Definition}
A schedule $\mathcal{S} = (\mathcal{R}, \mathcal{M}, TEC, TET)$ is defined as a tuple consisting of: A pool of resources ($\mathcal{R}$), a task to resource mapping ($\mathcal{M}$), total execution cost ($TEC$) and total execution time ($TET$). $\mathcal{M}$ is a mapping consisting of tuples of the form $(v_i, vm_{u_i}, ST_{v_i}, FT_{v_i})$, where task $v_i$ is allocated to resource $vm_{u_i}$ and has a start and end time of $ST_{v_i}$ and $FT_{v_i}$ respectively. $TEC$ includes both the computation and communication costs. The problem we address in this paper is that of minimizing the total cost subject to a deadline constraint ($D$), stated below. 
\begin{center}
    \textbf{Minimize:} $TEC$ \\
\textbf{Subject To:} $TET \leq D$
\end{center}

\section{Proposed Algorithm} \label{Sec:PA}
In this section, we describe our algorithm CEDCES, a PSO-based method. PSO was first proposed by Kennedy and Eberhart in 1995 \cite{kennedy1995particle} and is known to be quite effective in workflow scheduling \cite{pandey2010particle}, \cite{rodriguez2014deadline}, \cite{wang2019dynamic}, \cite{wu2010revised}. PSO contains a set of initial particles each corresponding to a particular solution to the problem at hand. The position and velocity for each particle $X_i^t$ are updated at each iteration depending on the particle's personal best ($pbest_i^t$) and the global best ($gbest^t$) till now, where t denotes the current iteration. The update equations are shown below:

\begin{equation} \label{eqn7}
    X_i^{t+1} = X_i^t + V_i^t
\end{equation}

\begin{equation} \label{eqn8}
    V_i^{t+1} = w \cdot V_i^t + c1 \cdot r1 \cdot (pbest_i^t - X_i^t) + c2 \cdot r2 \cdot (gbest^t - X_i^t)
\end{equation}
Here, $w$ and $c_1, c_2$ denote the inertia and acceleration coefficients respectively. $r_1$ and $r_2$ are two random numbers in between $0$ and $1$
 
\subsection{PSO Modeling}
\begin{itemize}
    \item \textbf{Encoding:} We follow the encoding used in \cite{rodriguez2014deadline}. A particle $X_i^t$ can be encoded as a vector of dimension $n = [u_1, u_2, \ldots,u_n]$, where each $u_i \in [\lvert \mathcal{R} \rvert]$ and it corresponds to the VM instance $vm_{u_i}$. As seen earlier, a client can lease any number of VM instances. So, to have the illusion of an unlimited set of resources we can consider one VM type for each task. However, this may lead to a very large resource pool making it difficult for PSO to converge. Instead, we consider $\mathcal{P}$ which is defined as the set having the maximum number of tasks that can be run in parallel. The pool of resources can now be set by considering one VM type for each task in $\mathcal{P}$. The size of the resource pool is given by Equation No. \ref{eqn9}
    
    \begin{equation} \label{eqn9}
        \lvert \mathcal{R} \rvert = \mathcal{N} \cdot \lvert \mathcal{P} \rvert
    \end{equation}
    where $\mathcal{N} = \sum_{k=1}^{m} \sum_{p=1}^{m_k} 1$ denotes the number of VMs across all the clouds. Computation of $\mathcal{P}$ is shown in Algorithm 1. As illustrated in Line No. 8 of Algorithm 1, we consider only tasks with $lvl < max\_lvl$. This is because for tasks with $lvl = max\_lvl + 1$, they will have at least one predecessor with $lvl = max\_lvl$ as seen from Equation No. \ref{eqn1} and all such tasks are already included in $\mathcal{P}$ in Line No. 4. Note that the obtained set may not be the largest one. \\
    While updating a particle's position using Equation No. \ref{eqn7}, we may encounter fractional values which are rounded to the nearest integer value. Further, it may happen that the particle goes out of bounds. For this we use the Shrink Method (SHR) \cite{alvarez2005mopso} that drags a particle back along its line of movement till it reaches the nearest boundary.
    
    \item \textbf{Fitness:} Since the fitness function is used to determine how good a particle is, it needs to reflect the objectives of the scheduling problem. Based on this, we set the fitness function to the total execution cost. However, the primary criterion for any particle is to satisfy the deadline constraint. While comparing two particles, the following cases might occur:
    \begin{itemize}
        \item Both particles satisfy the deadline constraint. In this case the particle with lesser execution cost is fitter.
        \item One particle satisfies the deadline constraint, while the other does not. In this case, the particle satisfying the constraint is considered fitter.
        \item Both particles violate the deadline constraint. In this case, the particle with lesser violation of constraint \emph{i.e.} lesser value of $TET$ is considered fitter.
    \end{itemize}
    The above constraint handling strategy is taken from \cite{deb2002fast}.

    \item \textbf{Particle to Schedule Mapping:} The pseudo-code to convert a particle to a schedule is shown in Algorithm 2, similar to the one in \cite{guo2018cost} and is described as follows. Initially, we compute the resource pool $\mathcal{R}$ by calculating the maximum number of tasks that can be executed in parallel. Also, we initialize the resources leased till now $\mathcal{R}_{curr}$ and task to resource mapping $\mathcal{M}$ to empty. For each task, we find the resource $vm_{u_i}$ it is mapped to and the corresponding VM type $VM(k, p)$ as shown in Line No. 5, 6. If a task has predecessors, it can start only after they finish as illustrated in Line No. 9, 10. For each successor node, we calculate the total transfer time and cost depending on where the tasks are hosted as shown from Line No. 12 to 21 using Equation No \ref{eqn3}. The processing time is computed in Line No. 22 using Equation No. \ref{eqn4}. If the resource is already available, we know that the task can start only after the current lease finish time $LFT_{vm_{u_i}}$ as shown in Line No. 23, 24. Otherwise, we first launch the corresponding VM instance with lease start time $LST_{vm_{u_i}}$ set as shown Line No. 27 and task start time after booting as shown in Line No. 26. Finally, the finish time of the task is calculated by adding the processing time to the start time as shown in Line No. 29. The lease finish time updated to the finish time of the task as shown in Line No. 30. After processing all the tasks, the total execution time is given by the finish time of the last task as shown in Line No. 32. Lines 33 to 35 compute the total execution cost.
    
    \item \textbf{Initialization:} Generally the particles for PSO are initialized with random positions and velocities \cite{guo2018cost}, \cite{rodriguez2014deadline}. Instead, we initialize the particle positions keeping in mind the deadline constraint and execution cost. Our initialization strategy is similar to that of \cite{sahni2015cost}. For this, we first define the notion of maximum execution time below.
    
    \begin{definition} \label{def6}
    $MET[v_i, VM(k, p)]$ denotes the maximum time it takes to execute a task $v_i$ up to $v_n$ along all possible paths, assuming the tasks along the path are executed on the same VM of type $VM(k, p)$. The formula for the same is presented in Equation No. \ref{eqn10} Note, as all tasks are executed on the same VM instance, we do not consider communication time.
    
    \begin{equation} \label{eqn10}
    \scriptsize
    MET[v_i, VM(k, p)] = \left\{ 
      \begin{array}{ c l }
      T_{exec}[v_i, VM(k, p)] & \quad \textrm{if } v_i = v_n \\
      \max_{v_j \in succ(v_i)} (MET[v_j, VM(k, p)]) + \\ T_{exec}[v_i, VM(k, p)] & \quad \textrm{otherwise}
      \end{array}
    \right.
    \end{equation}
    \end{definition}
    
    The pseudocode is shown in Algorithm 3. Our initialization scheme proceeds as follows. We order the tasks in such a way that a task always comes before any of its successors. The ordering of vertices can be done using topological sort \cite{kahn1962topological}. Random topological orders can be achieved by considering nodes with zero in-degree in a random order in the topological sort algorithm. Then we start assigning tasks to VMs in order. We compute the start time of task $v_i$ on VM instance $vm_r$ in Line No. 9-11. Since we are still determining where $v_i$ needs to be scheduled, the corresponding transfer time between predecessor $v_j$ and $v_i$ is not yet known. Hence we include it in the finish time of $v_j$ as shown in Line No. 11. Later, when the instance is determined, we update the finish time of $v_j$ as shown in Line No. 25. From the set of VMs which finish execution before the deadline (Line No. 14, 18), we choose the one with the least execution cost to schedule the current task in Line No. 21, 22. The corresponding lease period depends on whether the VM instance has been already launched or not as seen in Line No. 15, 19. Once the VM instance to schedule $v_i$ is determined, we first update the finish time of predecessor task $v_j$ and lease the finish time of its VM instances as shown in Line No. 25, 27 respectively. Since we do not know the transfer time between $v_i$ and its successors, we ignore the transfer time as shown in Line No. 34 and set the finish time and lease finish time of its VM instance in Line No. 35, 36 respectively. Finally, we assign the instance to $v_i$ in Line No. 37. Since we are not considering the transfer time while coming up with an allocation as mentioned earlier, the actual execution time may exceed the deadline. To avoid this, we choose a smaller deadline for Algorithm 3. In our experiments, we input a value of $0.9 \cdot D$ into the algorithm.
    
    \item \textbf{Crossover:} 
    We employ a random single-point crossover scheme in this paper. First two particles $X_i^t, X_j^t$ are chosen using a binary tournament selection mechanism \cite{miller1995genetic}. A random number $k \in [n]$ is chosen. The new particle is derived by considering $X_i^t$ from 1 to $k$ and appending it with $X_j^t$ from $k+1$ to n. After crossover, we replace the particle with the least fittest $pbest$ value with the new particle as done in \cite{chu1998genetic}.
    
    \item \textbf{Mutation:} 
    We employ a simple mutation scheme where a particle $X_i^t$ and index $k \in [n]$ are chosen at random. The value of $X_i^t$ $k$ is set to a random quantity $u_j \in [\lvert \mathcal{R} \rvert]$.

\end{itemize}

Having looked at all the components of PSO, we present our final procedure named CEDCES in Algorithm 4.

\subsection{Complexity Analysis}
\begin{itemize}
    \item \textbf{Algorithm 1:}
        Keeping track of the current level and the number of tasks per level during BFS gives us $max\_lvl$ and the number of tasks having $max\_lvl$. The time complexity of BFS is $\mathcal{O}(n^2)$. Sorting the tasks based on the topological level in list $L$ takes $\mathcal{O}(n \cdot \log n)$ time. For each task in list $L$, checking whether it shares dependency with any task already in $\mathcal{P}$ takes at most  $\mathcal{O}(n)$ time as at most n elements are present till now in $\mathcal{P}$. Hence the overall time complexity is $\mathcal{O}(n^2)$.
    \item \textbf{Algorithm 2:}
    Computing the order using topological sort takes $\mathcal{O}(n^2)$ time. Computing the maximum set of tasks takes $\mathcal{O}(n^2)$ time as seen from Algorithm 1. At maximum, there can be $\mathcal{N} \cdot n$ VM instances as seen from Equation No. \ref{eqn9}. Hence the total time to compute the resource pool will be $\mathcal{O}(n \cdot (\mathcal{N} + n))$. For each task (for loop Line No. 4), computing the start time and transfer time takes $\mathcal{O}(n)$ as seen from the for loops in Line No. 9-10 and 12-21 respectively. Hence overall the tasks, the total time required will be $\mathcal{O}(n^2)$. As discussed before, the resource pool $\mathcal{R}_{curr}$ can have a size of at most $\mathcal{O}(\mathcal{N} \cdot n)$. Hence the for loop in Line No. 33 will execute $\mathcal{O}(\mathcal{N} \cdot n)$ times. The body of the loop takes $\mathcal{O}(1)$ time. Hence, the overall time complexity will be $\mathcal{O}(n \cdot (\mathcal{N} + n))$.
    \item \textbf{Algorithm 3:}
    As seen in Algorithm 2, computing the resource pool takes $\mathcal{O}(n \cdot (\mathcal{N} + n))$ time. Computing the MET values for each task-VM pair takes $\mathcal{O}(n)$ time as seen from Equation No. \ref{eqn10}. Hence overall the task-VM pairs, $\mathcal{O}(\mathcal{N} \cdot n^2)$ time is required. For a particular task, for each VM instance pair (for loop at Line No. 6), computing the start time takes $\mathcal{O}(n)$ time as seen from Line No. 9-11. Rest of the statements take $\mathcal{O}(1)$ time. Overall VM instances, the total time complexity will be $\mathcal{O}(\mathcal{N} \cdot n^2)$. After finding the appropriate VM instance $vm_{u_i}$ for task $v_i$, rest of the update equations take $\mathcal{O}(n)$ time in Line No. 27-30 and 32-34. Overall tasks, the overall time complexity will be $\mathcal{O}(\mathcal{N} \cdot n^3)$.
    \item \textbf{Algorithm 4:}
    Initializing a particle takes  $\mathcal{O}(\mathcal{N} \cdot n^3)$ time as seen from Algorithm 3. Overall particles, $\mathcal{O}(NUM \cdot \mathcal{N} \cdot n^3)$ will be required. In each iteration, updating and repairing a particle takes $\mathcal{O}(n)$ time. Computing the schedule for each particle takes $\mathcal{O}(n \cdot (\mathcal{N} + n))$ as seen from Algorithm 2. The crossover and mutation schemes take $\mathcal{O}(n)$ time. Overall particles and iterations we get a time complexity of $\mathcal{O}(ITR \cdot NUM \cdot n \cdot (\mathcal{N} + n))$. The overall time complexity of the initialization and iteration phases is shown below: \\ $\mathcal{O}(NUM \cdot \mathcal{N} \cdot n^2 \cdot (n + ITR \cdot (\frac{1}{\mathcal{N}} + \frac{1}{n})))$.
\end{itemize}


\section{Experimental Evaluation} \label{Sec:Experiments}
In this section, we describe the experimental evaluation of the proposed solution approach. Initially, we describe the task graphs that we have used in this study.
\subsection{Task Graphs}
We have used the following benchmark task graphs in this study. These have been used in many other studies in the literature \cite{guo2018cost}, \cite{szabo2012evolving}, \cite{tang2021reliability}.
\begin{itemize}
    
    \item \textbf{Epigenomics}: Epigenomics applications are used to automate various operations in genome sequence processing. We consider task graphs with $n = 24, 100$ \cite{szabo2012evolving}.
    
    \item \textbf{LIGO}: LIGO applications are used to generate and analyze gravitational waveforms from data collected during the coalescing of compact binary systems. We consider task graphs with $n = 30, 100$ \cite{szabo2012evolving}.
    
\end{itemize}
The structure of each workflow can be obtained in XML format from the website \cite{graphgen} as described in \cite{bharathi2008characterization}. Due to space limitation, we only consider two workflows. More information about the workflows can be obtained from \cite{bharathi2008characterization}.

\subsection{Methods Compared}
We compare the performance of the proposed scheduling algorithm with the following state-of-art solutions from the literature:
\begin{itemize}
    \item \textbf{Particle Swarm Optimization (PSO)} \cite{rodriguez2014deadline}: This is a meta-heuristic cost-effective particle swarm optimization algorithm that aims to minimize the overall workflow execution cost while meeting deadline constraints.
    \item \textbf{Adaptive Particle Swarm Optimization with Genetic Algorithm (ADPSOGA)} \cite{guo2018cost}: This is a PSO-based method that aims to minimize the overall workflow execution cost while meeting deadline constraints in a multi-cloud environment. It employs a randomly two-point crossover operator and a randomly single-point mutation operator of the genetic algorithm. This method considers both computation and communication costs.
\end{itemize}
PSO considers workflow scheduling in a single cloud and ignores communication costs. As in \cite{guo2018cost}, it can be modified to include VMs from all cloud providers while encoding the solutions and considering communication costs.

\begin{figure*}[!ht]
\centering
\begin{tabular}{cccc}
\includegraphics[width=0.26\textwidth]{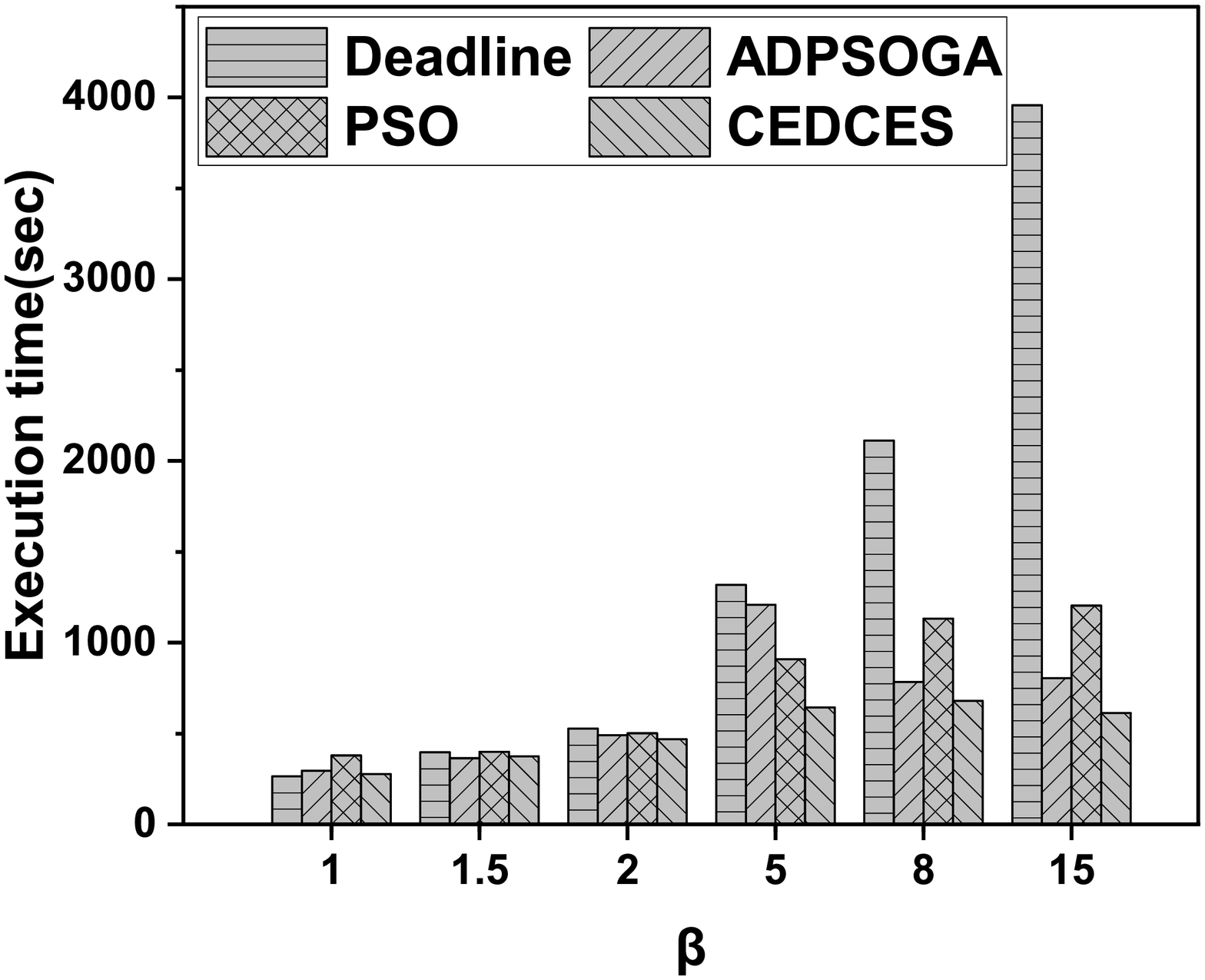} &
\includegraphics[width=0.26\textwidth]{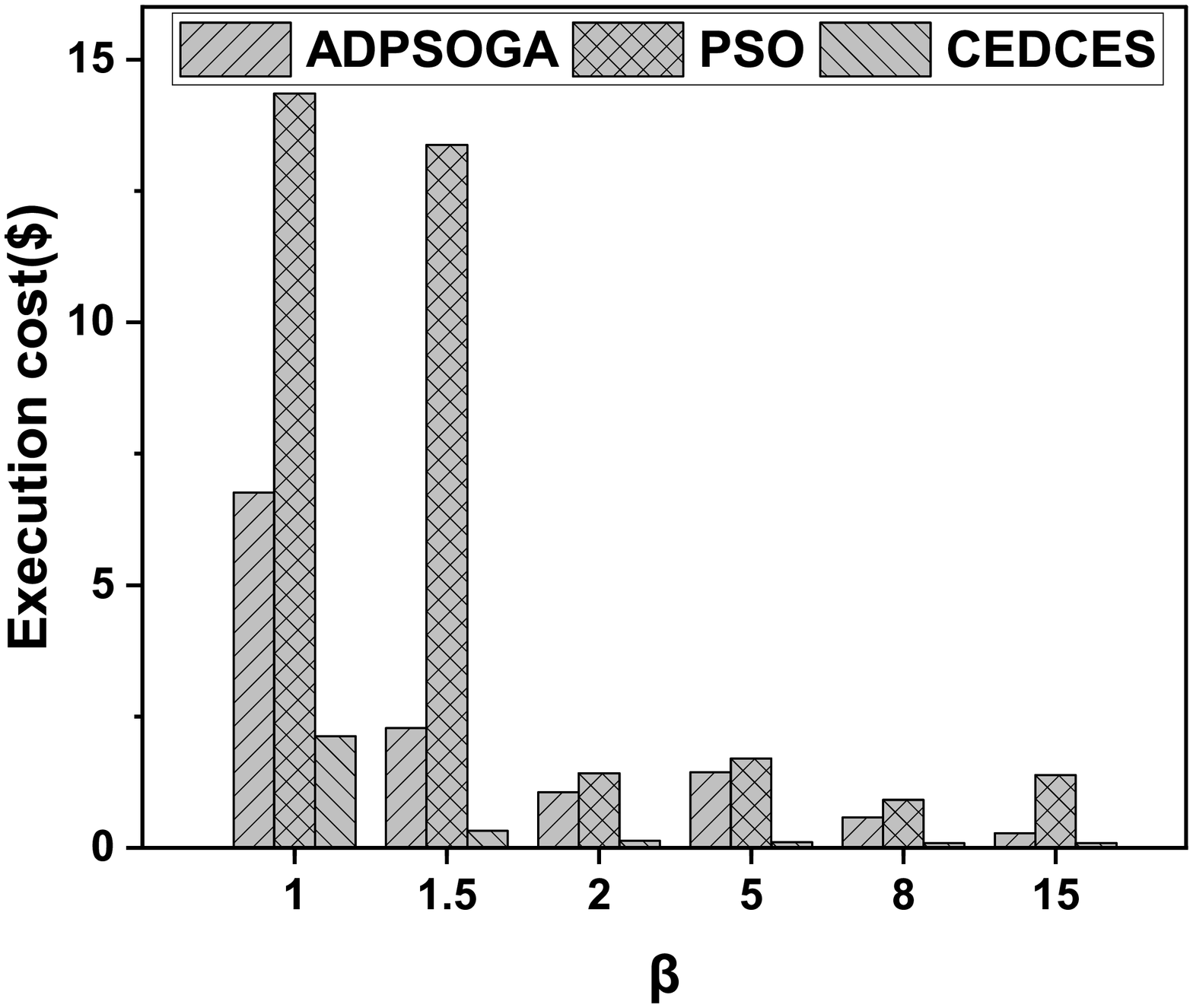} &
\includegraphics[width=0.26\textwidth]{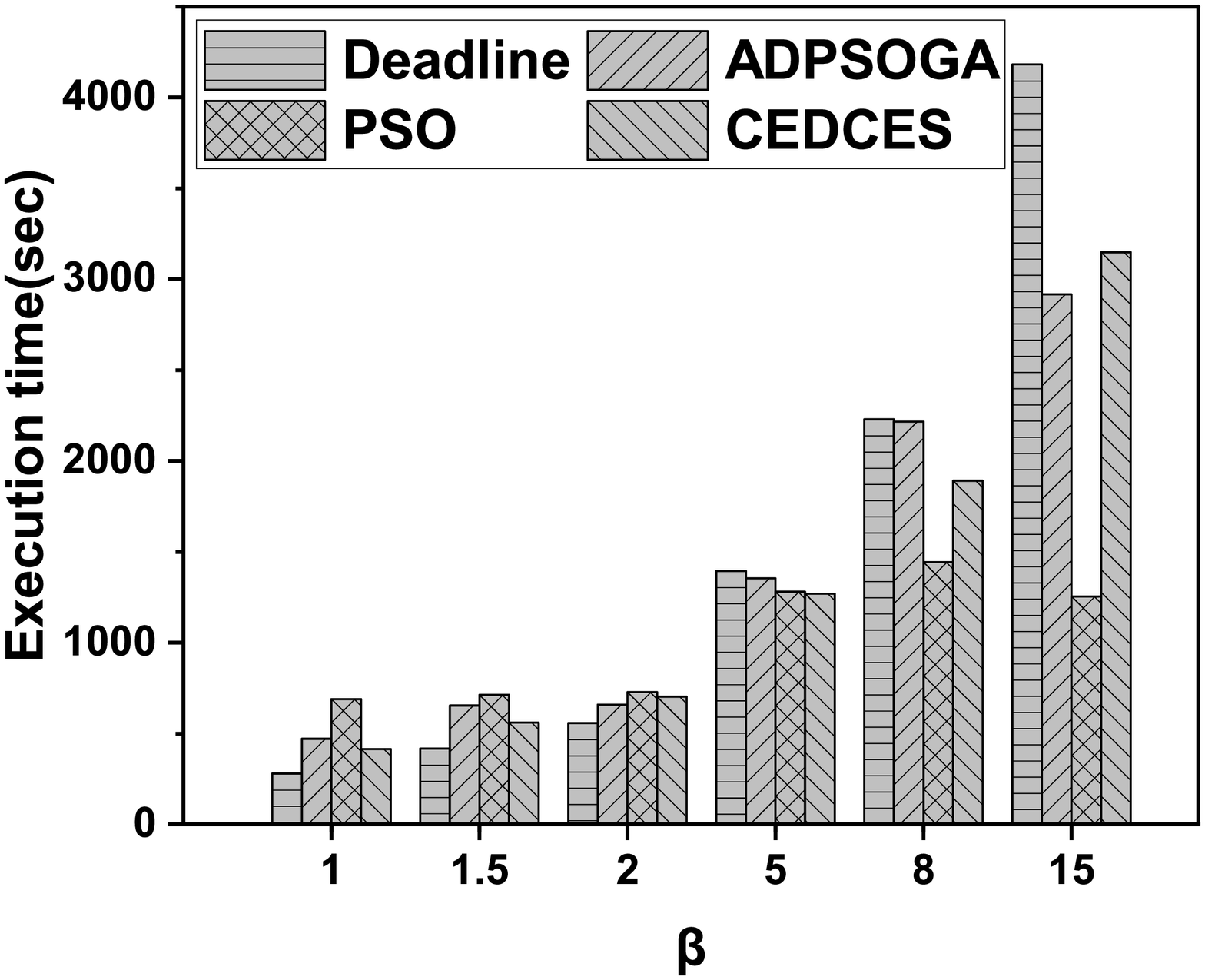} &
\includegraphics[width=0.26\textwidth]{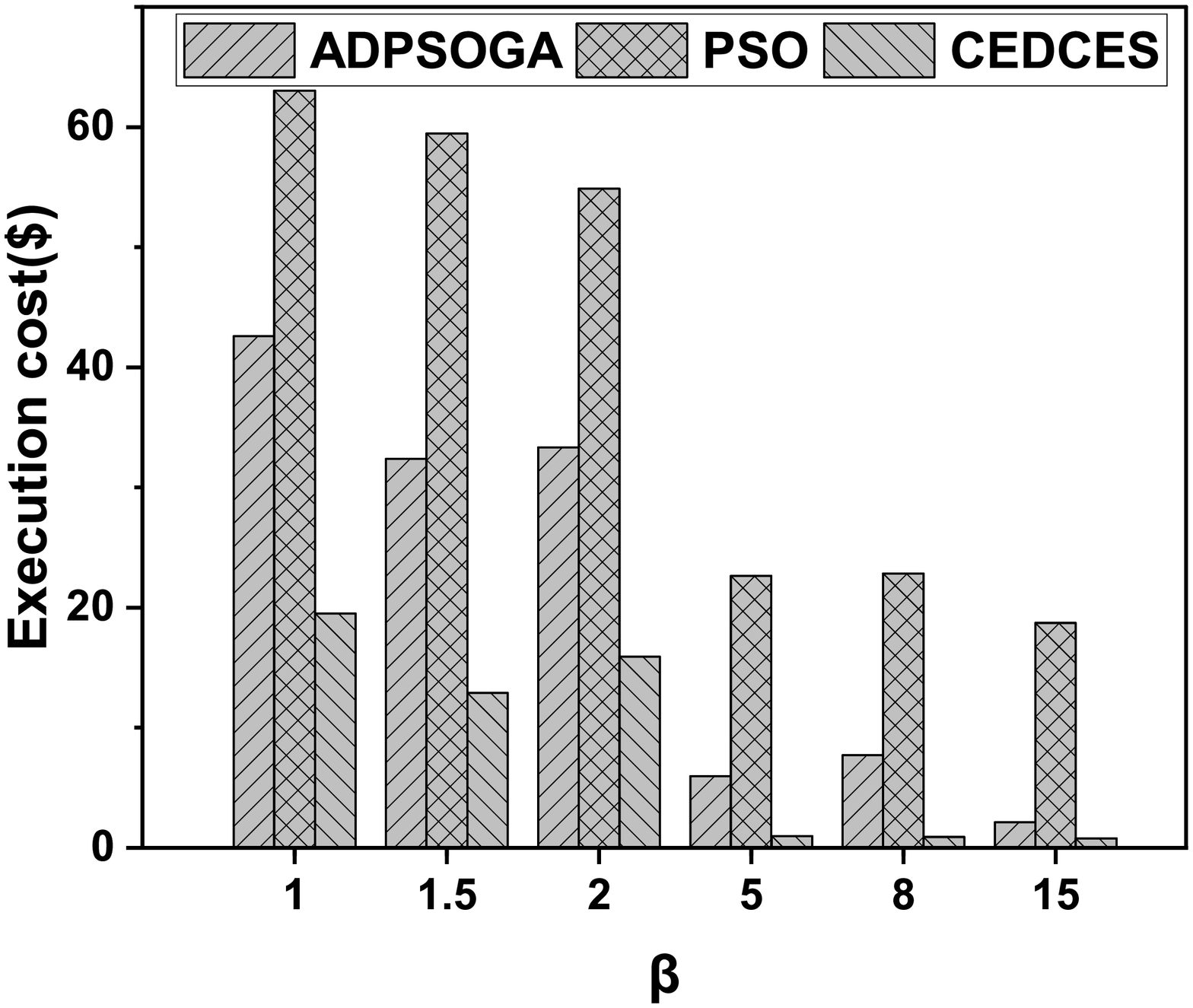} 
\\
(a) & (b) & (c) & (d) \\
\end{tabular}
\caption{Epigenomics workflows with n = 24, 100}
\label{fig1}
\end{figure*}

\begin{figure*}[!ht]
\centering
\begin{tabular}{cccc}
\includegraphics[width=0.26\textwidth]{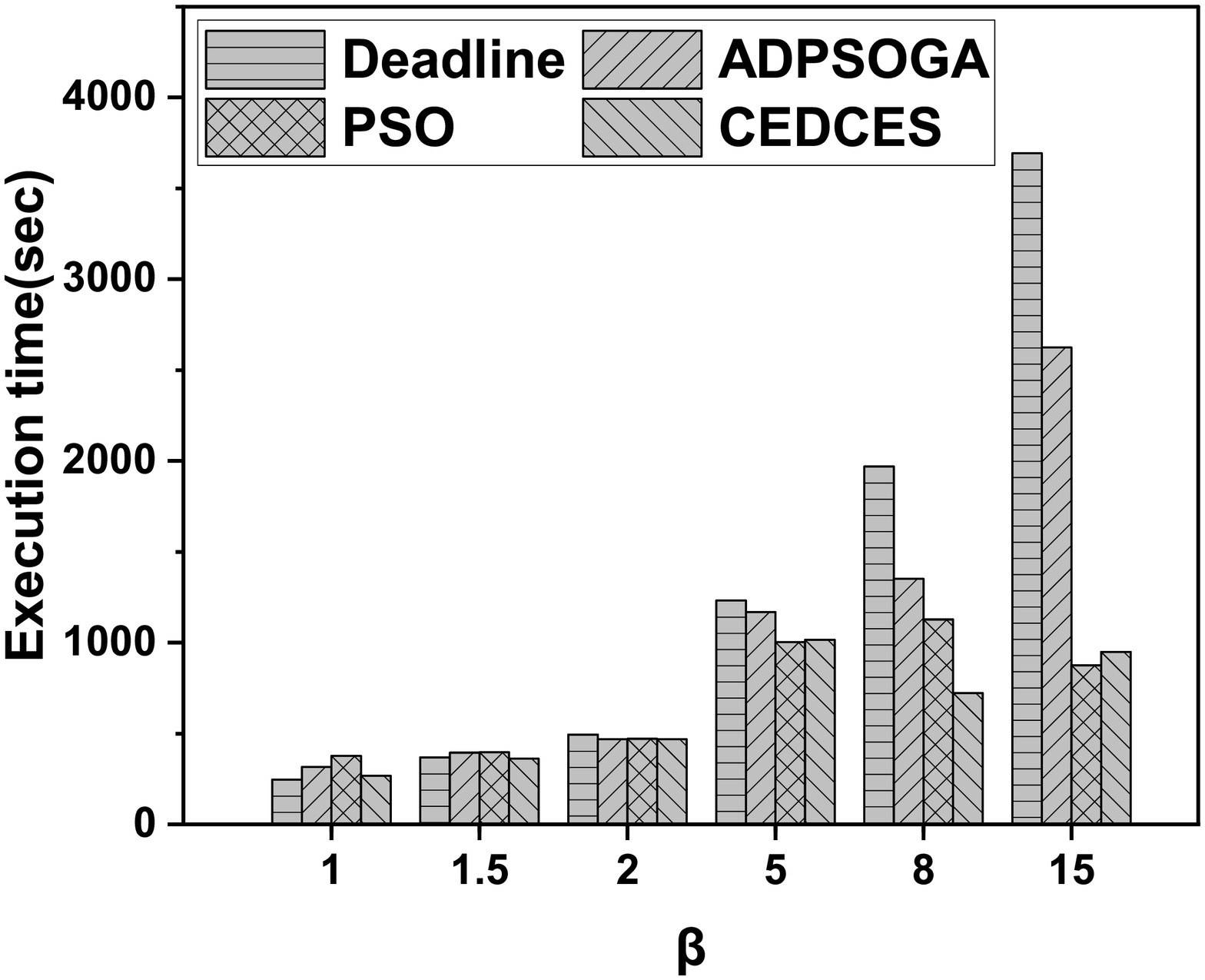} &
\includegraphics[width=0.26\textwidth]{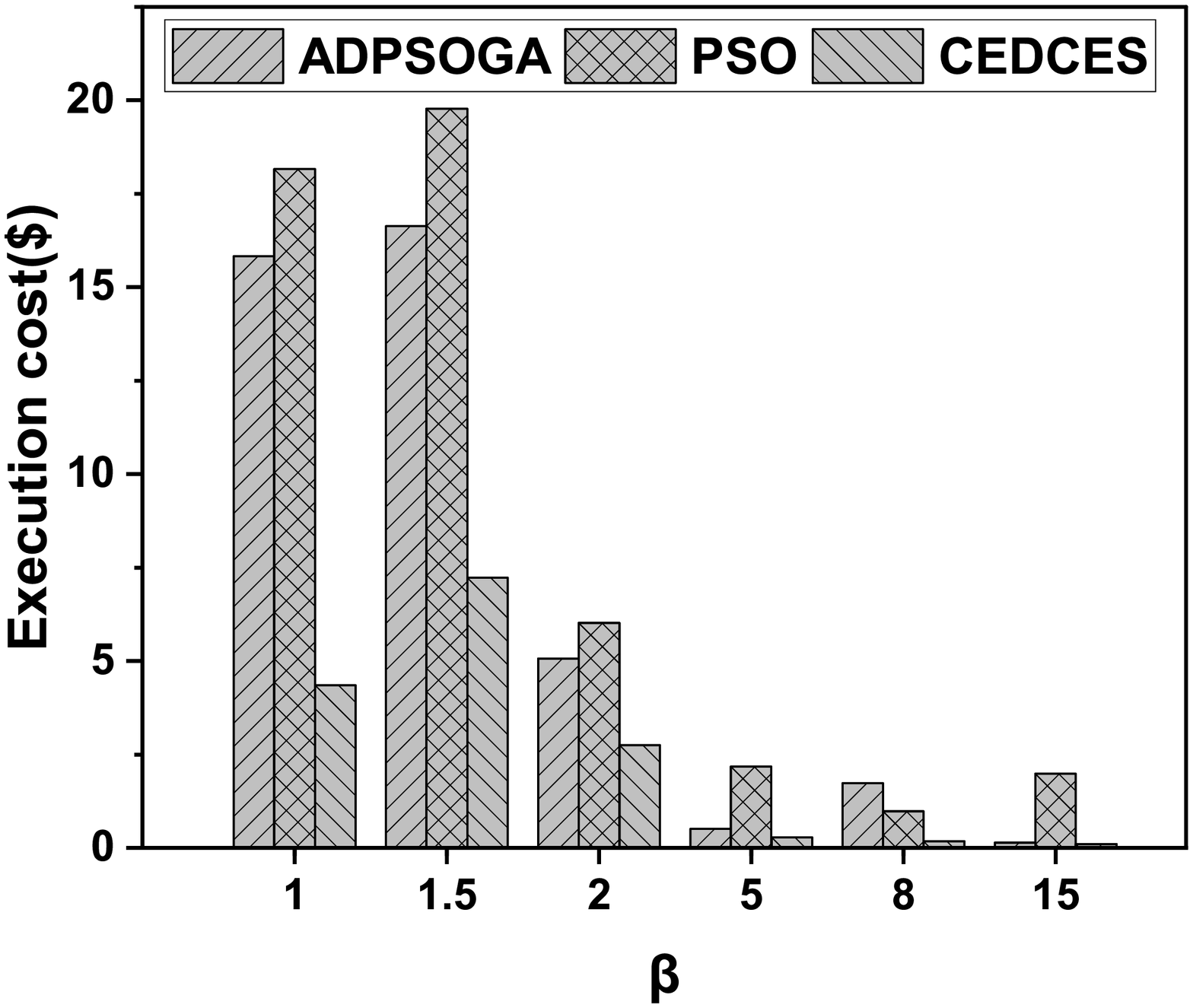} &
\includegraphics[width=0.26\textwidth]{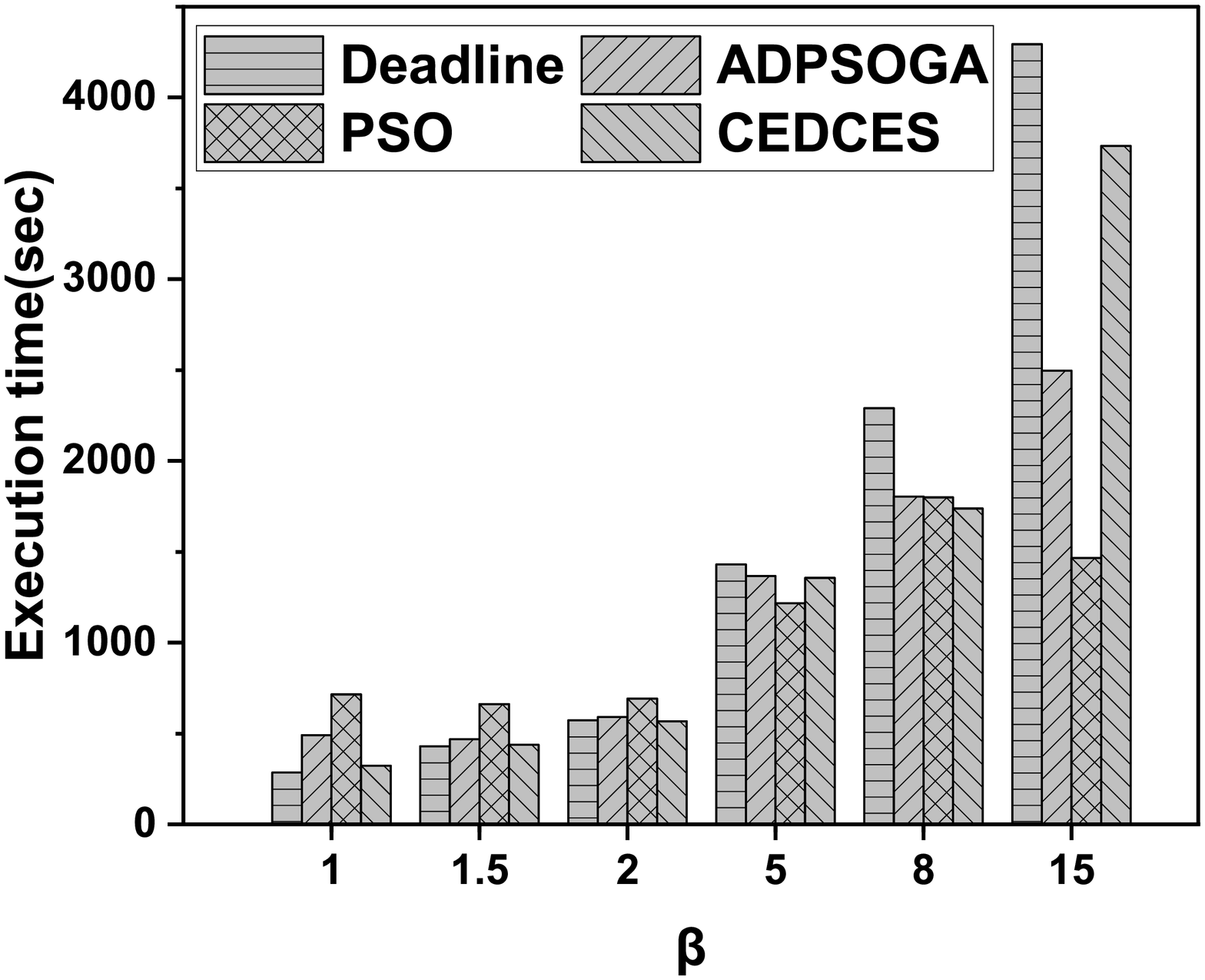} &
\includegraphics[width=0.26\textwidth]{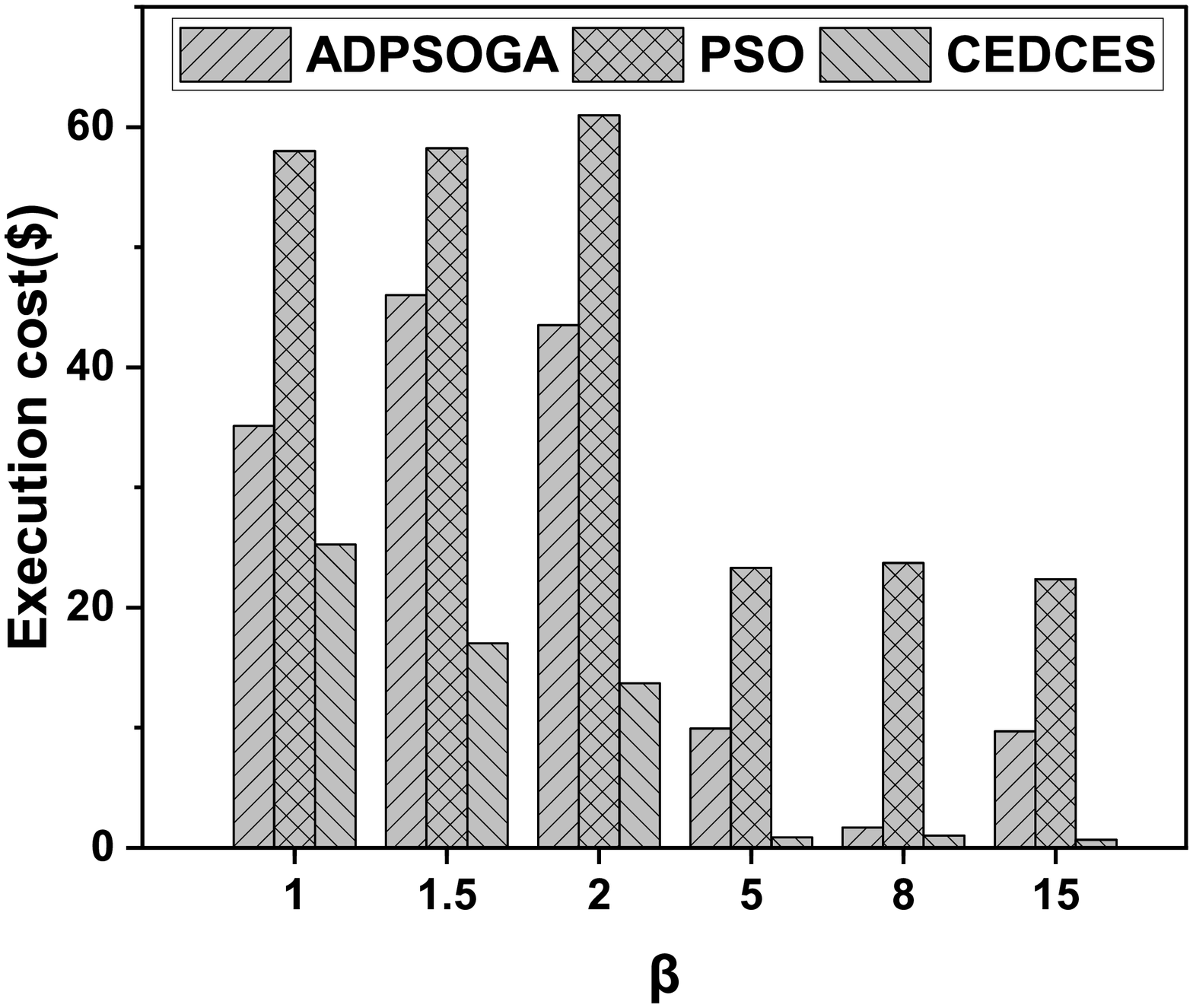} 
\\
(a) & (b) & (c) & (d) \\
\end{tabular}
\caption{LIGO workflows with n = 30, 100}
\label{fig2}
\end{figure*}

\subsection{Experimental Set Up and Description}
We implement the proposed solution approach on a workbench system having i5 10$^{\text{th}}$ generation processor and 32GB memory in Python 3.8.10. Parameters of our multi-cloud system are set as in \cite{tang2021reliability}. VMs are set with varying computation capacities in the range of 1 to 32. The average bandwidth internal to a cloud is set to 20Mbps and external bandwidth is set to 100Mbps. Pricing mechanisms are set proportional to VM compute capacities as in Table \ref{tab1} and data transfer costs according to Table \ref{tab2}. The boot time of each VM was set to 97 seconds \cite{rodriguez2014deadline}, \cite{guo2018cost}. The number of cloud providers considered is six, two from each type of \emph{MA}, \emph{AWS}, and \emph{GCP}. The two cloud providers having the same type are assumed to be located in different centers. The deadline constraint is set to $\beta \cdot Min(G)$ \cite{guo2018cost}, where $Min(G)$ is the makespan value of scheduling $G$ according to the HEFT method \cite{topcuoglu2002performance} and $\beta \in \{1, 1.5, 2, 5, 8, 15\}$. Low values of $\beta$ indicate tighter deadline constraints and higher values indicate looser constraints. Each experiment is run ten times and the average value is reported. In the end, all the algorithms are compared in terms of their execution cost and execution time.
\par The PSO parameters are set as follows: Each algorithm is with 100 particles for 1000 iterations as in \cite{guo2018cost},  $c1 = c2 = 2$. Finally, the weight parameter $w$ is set according to the tuning strategy as mentioned in \cite{shi1998modified} and also shown in Equation No. \ref{eqn11} where $w_{\max} = 1.4$ and $w_{\min} = 0.4$.

\begin{equation} \label{eqn11}
    w = w_{\max} - t \cdot \frac{(w_{\max} - w_{\min})}{ITR} 
\end{equation}

\subsection{Experimental Observations and Discussions}
\begin{itemize}
    \item \textbf{Epigenomics:} Fig. \ref{fig1} (a), (c) shows the plots of execution time and Fig. \ref{fig1} (b), (d) shows the plots of execution cost for n = 24 and 100, respectively. From Fig. \ref{fig1} (a), no algorithm satisfies the deadline constraint for $\beta=1$, but CEDCES achieves the least value having an overshoot of only 5.46\% compared to 43.47\% and 11.54\% for PSO and ADPSOGA,  respectively. For $\beta = 1.5$, only PSO does not satisfy the deadline having an overshoot of 1.21\%. For the remaining values of $\beta$, all the algorithms satisfy the deadline. From Fig. \ref{fig1} (b), CEDCES gives the least execution cost for all values of $\beta$ performing 91.34\% and 76.84\% better than PSO and ADPSOGA respectively.
    From  Fig. \ref{fig1} (c) we see that no algorithm satisfies the deadline for $\beta = $1, 1.5, and 2. For $\beta = 1, 1.5$, CEDCES achieves the least execution time having an overshoot of 48.62\%, 33.85\%. The same for PSO and ADPSOGA are 147.65\%, 70.45\%, and 68.85\%, 56.31\%, respectively. For $\beta = 2$, CEDCES gives overshoot of 25.89\% while PSO and ADPSOGA give 30.34\% and 18.24\%, respectively. For the remaining values of $\beta$, all the algorithms satisfy the deadline. From Fig. \ref{fig1} (d), CEDCES gives the least execution cost for all values of $\beta$ performing 78.88\% and 58.91\% better than PSO and ADPSOGA respectively.
    
     \item \textbf{LIGO:} Fig. \ref{fig2} (a), (c) shows the plots of execution time and Fig. \ref{fig2} (b), (d) shows the plots of execution cost for n = 30 and 100, respectively. From Fig. \ref{fig2} (a), no algorithm satisfies the deadline constraint for $\beta=1$, but CEDCES achieves the least value having an overshoot of only 9\% compared to 52.52\% and 28.2\% for PSO and ADPSOGA,  respectively. For $\beta = 1.5$, only CEDCES satisfies the deadline constraint while PSO and ADPSOGA have overshoot of 7.38\% and 6.5\% respectively. For the remaining values of $\beta$, all the algorithms satisfy the deadline. From Fig. \ref{fig2} (b), CEDCES gives the least execution cost for all values of $\beta$ performing 69.69\% and 62.71\% better than PSO and ADPSOGA respectively.
    From  Fig. \ref{fig2} (c) we see that no algorithm satisfies the deadline for $\beta = 1, 1.5$. For $\beta = 1, 1.5$, CEDCES achieves the least execution time having an overshoot of 12.94\%, 2.07\%. The same for PSO and ADPSOGA are 149.76\%, 54.23\%, and 71.26\%, 8.93\%, respectively. For $\beta = 2$, only CEDCES satisfies the deadline while PSO and ADPSOGA give 21.02\% and 2.93\%, respectively. For the remaining values of $\beta$, all the algorithms satisfy the deadline. From Fig. \ref{fig2} (d), CEDCES gives the least execution cost for all values of $\beta$ performing 76.25\% and 59.87\% better than PSO and ADPSOGA respectively.
\end{itemize}

In summary we observe that CEDCES gives least violation of deadline for low values of $\beta = 1, 1.5$ and least execution cost for all values of $\beta$.

\section{Conclusion and Future Research Directions} \label{Sec:CFA} 
In this paper, we have proposed a scheduling algorithm for the tasks of a workflow that aims to minimize the execution cost for a given deadline constraint. We integrate the VMs and pricing mechanisms offered by different cloud providers in our system's model. Formally we call the proposed solution approach as \textbf{C}ost \textbf{E}ffective \textbf{D}eadline \textbf{C}onstrained \textbf{E}volutionary \textbf{S}cheduler. The proposed approach is a modification of PSO including new initialization, crossover, and mutation schemes. We analyze the proposed solution approach to understand its time requirement. Experimental evaluation of the proposed solution approach with benchmark workflows has been done. We have observed that the proposed solution approach leads to much lesser execution cost or lesser deadline violation schedule compared to the existing solution approaches. Future studies on this problem can be done in the following ways. Firstly, more complex billing mechanisms can like classified computing fees, and reserved instances can be considered. Secondly, new methods for the initialization of the resource pool can be proposed. Also, the dynamic nature of the workflows can also be considered.

 \bibliographystyle{splncs04}
 \bibliography{Paper}

\begin{thebibliography}{10}
\providecommand{\url}[1]{\texttt{#1}}
\providecommand{\urlprefix}{URL }
\providecommand{\doi}[1]{https://doi.org/#1}

\bibitem{alvarez2005mopso}
Alvarez-Benitez, J.E., Everson, R.M., Fieldsend, J.E.: A mopso algorithm based
  exclusively on pareto dominance concepts. In: International conference on
  evolutionary multi-criterion optimization. pp. 459--473. Springer (2005)

\bibitem{bharathi2008characterization}
Bharathi, S., Chervenak, A., Deelman, E., Mehta, G., Su, M.H., Vahi, K.:
  Characterization of scientific workflows. In: 2008 third workshop on
  workflows in support of large-scale science. pp. 1--10. IEEE (2008)

\bibitem{bundy1984breadth}
Bundy, A., Wallen, L.: Breadth-first search. In: Catalogue of artificial
  intelligence tools, pp. 13--13. Springer (1984)

\bibitem{chu1998genetic}
Chu, P.C., Beasley, J.E.: A genetic algorithm for the multidimensional knapsack
  problem. Journal of heuristics  \textbf{4}(1),  63--86 (1998)

\bibitem{aws}
data~transfer costs, A.W.S.: [online].
  \url{https://aws.amazon.com/ec2/pricing/on-demand/} (Accessed 11 September
  2022)

\bibitem{gcp}
data~transfer costs, G.C.P.: [online].
  \url{https://cloud.google.com/vpc/network-pricing} (Accessed 11 September
  2022)

\bibitem{ma}
data~transfer costs, M.A.: [online].
  \url{https://azure.microsoft.com/en-in/pricing/details/bandwidth/} (Accessed
  11 September 2022)

\bibitem{deb2002fast}
Deb, K., Pratap, A., Agarwal, S., Meyarivan, T.: A fast and elitist
  multiobjective genetic algorithm: Nsga-ii. IEEE transactions on evolutionary
  computation  \textbf{6}(2),  182--197 (2002)

\bibitem{graphgen}
Generator, W.: [online].
  \url{https://confluence.pegasus.isi.edu/display/pegasus/Deprecated+Workflow+Generator}
  (Accessed 14 September 2022)

\bibitem{guo2018cost}
Guo, W., Lin, B., Chen, G., Chen, Y., Liang, F.: Cost-driven scheduling for
  deadline-based workflow across multiple clouds. IEEE Transactions on Network
  and Service Management  \textbf{15}(4),  1571--1585 (2018)

\bibitem{kahn1962topological}
Kahn, A.B.: Topological sorting of large networks. Communications of the ACM
  \textbf{5}(11),  558--562 (1962)

\bibitem{kennedy1995particle}
Kennedy, J., Eberhart, R.: Particle swarm optimization. In: Proceedings of
  ICNN'95-international conference on neural networks. vol.~4, pp. 1942--1948.
  IEEE (1995)

\bibitem{miller1995genetic}
Miller, B.L., Goldberg, D.E., et~al.: Genetic algorithms, tournament selection,
  and the effects of noise. Complex systems  \textbf{9}(3),  193--212 (1995)

\bibitem{pandey2010particle}
Pandey, S., Wu, L., Guru, S.M., Buyya, R.: A particle swarm optimization-based
  heuristic for scheduling workflow applications in cloud computing
  environments. In: 2010 24th IEEE international conference on advanced
  information networking and applications. pp. 400--407. IEEE (2010)

\bibitem{rodriguez2014deadline}
Rodriguez, M.A., Buyya, R.: Deadline based resource provisioningand scheduling
  algorithm for scientific workflows on clouds. IEEE transactions on cloud
  computing  \textbf{2}(2),  222--235 (2014)

\bibitem{roy2020contention}
Roy, S.K., Devaraj, R., Sarkar, A., Maji, K., Sinha, S.: Contention-aware
  optimal scheduling of real-time precedence-constrained task graphs on
  heterogeneous distributed systems. Journal of Systems Architecture
  \textbf{105},  101706 (2020)

\bibitem{sahni2015cost}
Sahni, J., Vidyarthi, D.P.: A cost-effective deadline-constrained dynamic
  scheduling algorithm for scientific workflows in a cloud environment. IEEE
  Transactions on Cloud Computing  \textbf{6}(1),  2--18 (2015)

\bibitem{shi1998modified}
Shi, Y., Eberhart, R.: A modified particle swarm optimizer. In: 1998 IEEE
  international conference on evolutionary computation proceedings. IEEE world
  congress on computational intelligence (Cat. No. 98TH8360). pp. 69--73. IEEE
  (1998)

\bibitem{szabo2012evolving}
Szabo, C., Kroeger, T.: Evolving multi-objective strategies for task allocation
  of scientific workflows on public clouds. In: 2012 IEEE Congress on
  Evolutionary Computation. pp.~1--8. IEEE (2012)

\bibitem{tang2021reliability}
Tang, X.: Reliability-aware cost-efficient scientific workflows scheduling
  strategy on multi-cloud systems. IEEE Transactions on Cloud Computing  (2021)

\bibitem{topcuoglu2002performance}
Topcuoglu, H., Hariri, S., Wu, M.Y.: Performance-effective and low-complexity
  task scheduling for heterogeneous computing. IEEE transactions on parallel
  and distributed systems  \textbf{13}(3),  260--274 (2002)

\bibitem{ullman1975np}
Ullman, J.D.: Np-complete scheduling problems. Journal of Computer and System
  sciences  \textbf{10}(3),  384--393 (1975)

\bibitem{wang2019dynamic}
Wang, Z.J., Zhan, Z.H., Yu, W.J., Lin, Y., Zhang, J., Gu, T.L., Zhang, J.:
  Dynamic group learning distributed particle swarm optimization for
  large-scale optimization and its application in cloud workflow scheduling.
  IEEE transactions on cybernetics  \textbf{50}(6),  2715--2729 (2019)

\bibitem{wu2010revised}
Wu, Z., Ni, Z., Gu, L., Liu, X.: A revised discrete particle swarm optimization
  for cloud workflow scheduling. In: 2010 international conference on
  computational intelligence and security. pp. 184--188. IEEE (2010)

\end{thebibliography}
\end{document}